# Change Blindness in 3D Virtual Reality

Madis Vasser[1]*, Markus Kängsepp[2], Jaan Aru[2,3].

[1] Institute of Psychology, University of Tartu, Tartu, Estonia

[2] Institute of Computer Science, University of Tartu, Tartu, Estonia

[3] Institute of Public Law, University of Tartu, Tartu, Estonia

*corresponding author: Madis Vasser

e-mail: madis.vasser@ut.ee

## Abstract

In the present change blindness study subjects explored stereoscopic three dimensional (3D) environments through a virtual reality (VR) headset. A novel method that tracked the subjects' head movements was used for inducing changes in the scene whenever the changing object was out of the field of view. The effect of change location (foreground or background in 3D depth) on change blindness was investigated. Two experiments were conducted, one in the lab (n = 50) and the other online (n = 25). Up to 25% of the changes were undetected and the mean overall search time was 27 seconds in the lab study. Results indicated significantly lower change detection success and more change cycles if the changes occurred in the background, with no differences in overall search times. The results confirm findings from previous studies and extend them to 3D environments. The study also demonstrates the feasibility of online VR experiments.

**Introduction**

Change blindness occurs when a person is unable to notice a big change in a scene after a visual disruption, unless the scenario is presented numerous times or a hint is given (Simons & Rensink, 2005). Change blindness literature gives valuable insights about the processes and limitations of human attention. Previous works with unfamiliar 2D scenes show that changes in the foreground are easier to detect (Mazza et al, 2005; Turatto et al, 2002) and the sudden appearance of an object tends to be more prominent than disappearance (Cole & Liversedge, 2006). Although the role of situational context has received little study in attention research (Smilek et al, 2006), marginal interest changes in the foreground seem to be harder to detect than central changes (O'Regan et al, 2000). Robust change blindness can occur even in familiar and well known mental scenes (Rosielle & Scaggs, 2008).

A typical change blindness protocol consists of a stimulus scene, an altered version of the same scene (both shown around 500ms) and a short (100-200ms) visual distraction in between to mask the change. A more exploratory approach has sometimes been used, where the subjects' own behaviour determines the moment of the change (O'Regan et al, 2000; Suma et al, 2011). Eye movements, blinks, transient masks ("flicker paradigm") or partial patterns ("mud splashes") have been used previously for introducing the change (Simons & Rensink, 2005). Different types of changes have been explored, including the abrupt appearance of a new object, the sudden disappearance of an existing object, sudden or gradual changes in color and shape, position and movement (Karacan, 2010). Various parameters affect the total search time, such as the number of objects in the scene, their overall placement (random or in a pattern), colour, shape and how probable the change is (Gusev & Mikhaylova, 2013).

A valuable way to study attention is allowing participants to explore their environment (Scarfe & Glennerster, 2015; Smilek et al, 2006). With traditional display systems the sensory input is merely audio-visual and there are no means to interact with the reality presented on a passive screen (Pillai, Schmidt & Richir, 2013). Some investigators have sought to more ecologically valid experimental conditions (Smilek et al, 2006) or actually performed the experiments as field studies (Simons & Levin, 1998). As real natural environments are difficult to control experimentally and reproduce, and 2D images are lacking many real-life features, 3D virtual reality (VR) environments are sometimes preferred for cognitive psychology research (Karacan, 2010). VR environments render quasi-realistic natural scenes, giving the experimenter absolute control over all details of the scene, and allow perfect reproduction of the experimental setting between subjects and studies (Triesch et al, 2003). This approach also gives more freedom of movement to the study subject, who is not confined to look only in a single direction. This is important, as maximizing strict control over the subject's behaviour might not reveal important aspects of complex systems such as human cognition and attention in the real world (Smilek et al, 2006).

VR systems of today are capable of high level of immersion and the feeling of presence. Here, presence refers to the perception of one's surrounding as mediated by both automatic and controlled mental processes, an experience of a different reality (Pillai, Schmidt & Richir, 2013). High level of presence in a virtual study environment might yield stronger cognitive ethology (Smilek et al, 2006), resembling the high perceptual and computational demands present in real life behaviors (Scarfe & Glennerster, 2015; Shinoda, Hayhoe, & Shrivastava, 2001).

The present paper uses quasi-realistic VR scenes and a novel method of introducing changes in the scenes. Our aim is to study the effect of actual spatial distance of the change as a variable of change blindness. With traditional 2D displays this has been unfeasible. The changes in the present study occur whenever the subject has their head turned completely away from the changing object. This method does not use a visual transients like the flicker or mud splash paradigms, ensuring a more comfortable and natural visual experience for the study subject. Based on the literature on foreground and background effects on change blindness in 2D scenes (Mazza et al, 2005; Turatto et al, 2002) and attentional blanks stares and attentional dead zones (Caplovitz et al, 2008; Utochkin, 2011), we predict that changes occurring in the foreground are significantly easier to detect than changes in the background.

## EXPERIMENT 1

In the first experiment we studied the effects of change distance to the observer on change blindness performance, using an offline laboratory protocol.

### Materials and methods

### Participants

50 study subjects (mean age 24 years, SD = 3.8,  equal number of males and females) participated in the experiment. The sample size was determined by previous studies (Suma et al, 2011, Triesch et al, 2003). All reported having normal or corrected to normal vision. 11 subjects in the final analysis had to remove their prescription glasses in order to participate in the experiment. Before the experiment all subjects gave written informed consent. Some participants received course credit. The experiments were undertaken in compliance with national legislation and the Declaration of Helsinki.

**Stimuli and procedure**

Twelve quasi-realistic 3D scenes of a typical livingroom setting were used in the experiment (figure 1 A). Three additional scenes in the beginning of the experiment were used to familiarize participants with the VR headset and study methodology. The 12 scenes used in the experimental block were balanced so that 6 rooms had changes occurring in the foreground condition (1-3 meters from the observer in virtual space) and the other 6 rooms in the background condition (4-6 meters from the observer). These arbitrary distances were chosen from an earlier small pilot study. There was also an equal distribution of changes in the middle of the room and in the periphery. Every room began with the location of the change in the field of view. All changes between the conditions were approximately equated with respect to their visual size and contrast (see figure S1 in the Supplemental Material available online for all the actual changes).

(A)                                                (B)

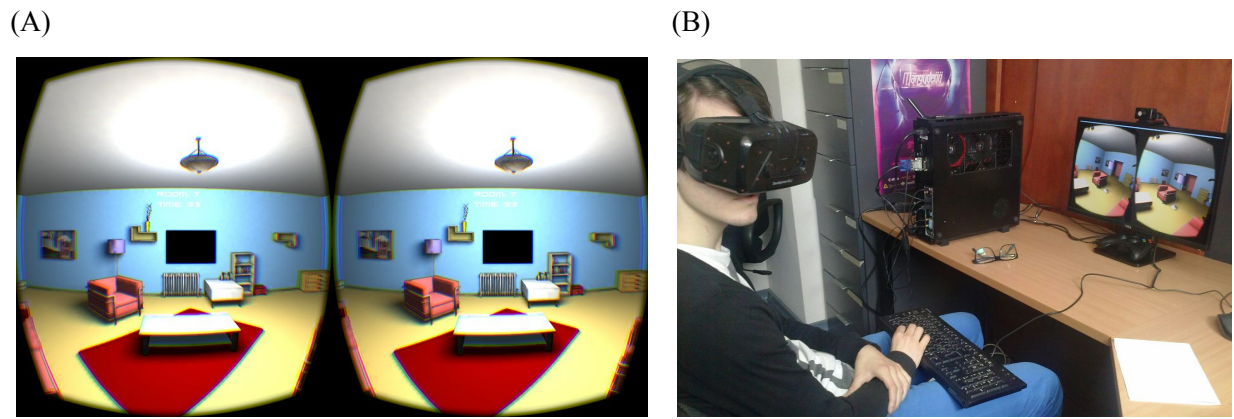

Figure 1. (A) A typical livingroom setting used in the experiment, as seen through the stereoscopic VR headset. Images for the left and right eye, accordingly. Image warping and chromatic distortion is introduced to produce the correct output through the lenses of the VR headset. (B) A study subject with the virtual reality goggles, headphones and a keyboard for responding.

The general category of all the changing objects were held constant between different conditions (e.g., furniture). All the changing objects were central (in a prominent place), context relevant (logical part of the interior), probable (easily movable if compared to real-world analogues), same colour (between foreground and background conditions). Every room contained the same number of objects in different configurations. The structural properties of the rooms were constant. Therefore, proximity of the changing object to the observer was the only parameter manipulated. The size of the changing objects was chosen to produce a comparable retinal image between foreground and background conditions. However, since our setup also allowed the study subjects to lean approximately 10 centimeters in every direction in virtual space, the size of the retinal image varied slightly. The changing objects cycled between visible and not visible states whenever the object was out of the field of view of the headset. Many small pilot-studies were conducted to find the optimal placement of the changing objects and to refine the study instructions given to the subjects.

The experiment was introduced to the subjects as a study of perception and attention. Participants were instructed to actively search for one constantly changing object in every new scene. They were explicitly told that the change would occur while they were looking elsewhere. Subjects were encouraged to press the response key as soon as they were certain they had spotted the change, after which the timer would stop and they had the chance to mark the object and confirm the selection. After that the next scene would begin. To facilitate active monitoring of the surroundings and to prevent stalling, a time limit of 60 seconds was set for each scene. Participants sat on a static chair, wearing a VR headset and headphones (for sound isolation) and holding a keyboard on their lap (for a typical setup see figure 1 B).

At the beginning of the experiment participants completed three practice trials with verbal and visual instructions, followed by 12 experimental trials. The 12 experimental trials included six trials with a foreground change and six trials with a background change. Experimental trials were presented in random order to minimize further practice effect on the results. After the change blindness task the subject filled a short questionnaire about gender, age, eyesight and other items (table 1). The experiment lasted approximately 20 minutes.

**Apparatus**

The 3D environments were constructed using Unity 4.6 game engine with the help of a custom virtual reality toolbox specifically designed for this experiment (Kängsepp, 2015; Vasser et al, 2015). The following data was automatically collected for all participants in every trial: the answer (true or false), number of times the change took place (from visible to invisible and vice versa), search time, pause time (when giving the answer) and rotational head movement data. For 16 participants the time intervals of the change cycles were also collected. The program was presented to the study subjects using the Oculus Rift Development Kit 2 virtual reality headset (Oculus VR, LLC) with a low persistence OLED display, 100 degree field of view, 75hz refresh rate and 960 x 1080 pixel resolution per eye. The system was running on a pc with Intel Core i7-4970K, MSI GeForce GTX 970 OC and 8GB of RAM.

**Data cleaning and analysis**

Of the total subject pool of 50 participants, 46 subjects were included in the final experimental data analysis. Three subjects were excluded due to a high number (>2) of answering errors caused by misinterpretations of the instructions. One subject stopped the experiment half-way due to personal discomfort. From the remaining 552 trials, one trial was completely omitted due to corrupted data. Two of the remaining 551 trials lacked search time data due to technical reasons. Since one subject missed all of the changes in the background conditions, for some statistical tests these trials were left out of the analysis. The main within-subject independent variable in the experiment was change location, which included two conditions: foreground or background. Dependent variables were as follows: missed changes (out of time or a false answer), mean search time in seconds and mean number of change cycles (for trials with successful detection). From 16 subjects the time intervals between the changes were recorded. From a total of 1013 intervals 9 were removed for being over 30 seconds long - this was over half of the control time in every room and a clear sign of an outlier.

The analysis was conducted in R (R Core Team, 2013). The distributions of search times and proportion of missed changes were probed for deviations from the normal distributions with the Shapiro–Wilk test. If there were deviations, the Wilcoxon signed-rank test was used instead of the t-test. All statistical analyses were only done once, after the data was collected.

**Results**

The effects of change blindness were large in the present study. Subjects completely failed to identify or misidentified the changed objects in 137 of the 551 trials (24.9%). For all the successful trials, the mean search time was 26.9 seconds (SD = 14.4) and on average, the changing object changed states 4.7 times (SD = 3.9) before being spotted. Only 6.5% of study subjects (3 out of 46) managed to successfully spot all the 12 changes presented in the experiment.

First the number of errors (false answer or out of time trials) was examined between conditions from the sample of 551 trials. On the proportion of trials where the person detected a change, with a minimum of 0 (no trials with successful detection) and the maximum of 1 (all trials with successful detection), the foreground condition yielded a mean value  of 0.79 (SD = 0.17) and the background condition 0.71 (SD = 0.23). A two-tailed paired t-test assuming unequal variances revealed a statistically significant difference between the proportions of successful trials (t = 2.43, p = 0.019. d = 0.39). Changing objects farther away from the subject were detected less successfully as compared to the changing objects closer to the subject. All following analysis were conducted without the data from the error trials.

When comparing the amount of changes needed for successful detection on 414 trials, a significant difference was found using Wilcoxon signed rank test with continuity correction (V = 269, p = 0.037. d = 0.37). The mean amount of changes for the foreground and background conditions were 4.27 (SD = 1.86) and 5.02 (SD = 2.12), respectively. There had to be more changes for background objects so that subjects would notice them.

Search time analysis was performed on 412 trials. The mean search time for the foreground condition was 27.56 seconds (SD = 8.64) and for the background condition 27.95 seconds (SD = 8.47). The difference between the means was not statistically significant (t = -0.38, p = 0.71).  Results between conditions are shown on figure 2 A.

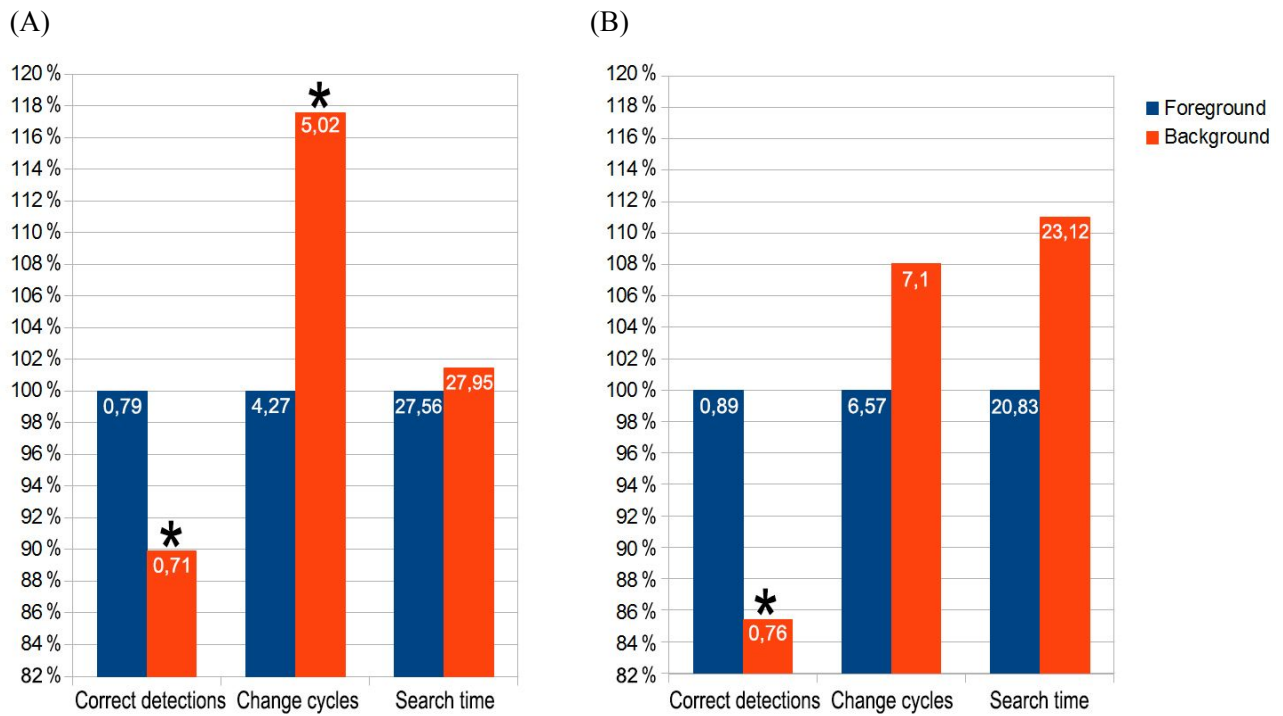

Figure 2. Summary data for the foreground and background conditions from experiment 1 (A) and experiment 2 (B) as percentage of the foreground condition values (y axis). Numbers on the bars show absolute values. Asterisks denote statistically significant differences (p < 0.05). Note that we used the percentage scores to have all three variables (proportion of correct detections, amount of change cycles and search time) on the same axis.

Table 1

*Background questions answered by the participants in Experiment 1*

| | |
|---|---|
| Prior experience with computer games (1-5) | M= 3.24, SD = 1.37 |
| Nausea/dizziness during the experiment (1-5) | M=1.2, SD = 0.6 |
| Did you try to remember to objects in the room (yes) | 95.7% |
| Did you use any special mnemonic techniques (yes) | 10.9% |
| Did you check the objects one by one (yes) | 56.5% |
| Did you move your head repeatedly to spot the changes (yes) | 56.5% |
| Did you notice the difference of the shadows (yes) | 8.7% |
| Did you use the room plan as a hint (yes) | 47.7% |
| Did you use the experimental plan as a hint (yes) | 26.1% |
| Did you complete the experiment without any strategies (yes) | 0.0% |

Subjects exhibited and reported different strategies for completing the experimental task (table 1). To study the effects of strategies used by at least 20% of the participants, we ran separate ANOVAs with the factors experimental condition (foreground vs background) and strategy (whether the person used or did not use one of the above mentioned strategies). Comparing the effect of different strategies on the amount of changes needed for a successful detection, turning one's head repeatedly was shown to significantly increase the amount of required change cycles (F = 15.20, df = 43, p = 0.00033, generalized eta squared (ges) = 0.20). This difference was more pronounced in the background condition as evidenced by the interaction between this strategy and the experimental condition (F = 5.32, df = 43, p = 0.026, ges = 0.037). This result leads to the obvious suspicion that one of the the main findings (see above) showing that the the amount of changes before successful detection is higher for background objects could be mainly driven by subjects who used such repeated head movement strategy. Indeed, when the amount of changes needed was analysed only in the subgroup  (n = 20) that did not use repeated head movements, no effect of distance was observed on the amount of changes needed (p > 0.3). However, as expected, the effect of distance was strong within the group of subjects (n = 26) who used the strategy of repeated head movements. In this group, the detection of background objects required more changes (V = 56, p = 0.01, d

= 0.61). Repeated head movements was also the only strategy that led to significantly shorter detection times in the foreground condition as evidenced by an interaction between the strategy and the experimental condition for search time in successful trials (F = 5.59, df = 43, p = 0.022, ges = 0.031). The subjects who used repeated head movements spotted the foreground objects quicker (t = -2.29, p = 0.028. d = 0.71) with an average of 5.82 seconds less required to spot the change in the foreground compared to the group without this strategy. Importantly, repeated head movements did not influence the effect of the experimental condition on the proportion of trials where the person detected a change (interaction between the strategy and experimental condition F < 1)

Differences between males and females were analyzed. The groups did not differ in change amounts (t = -1.52, p = 0.136) but a significant difference was found in search time (t = -2.31, p = 0.03. d = 0.68). The average for females was 30.32 seconds (SD = 6.56), for males it was 25.45 seconds (SD = 7.67). Males were quicker to spot the changing object.

After the experiment, subjects also assessed their prior experiences with computer games on a subjective 5-point scale (mean 3.24, SD = 1.37). The subjects were split into two groups to analyze the effects of computer game experience on change blindness, with the split point of the scale being at 3. The experienced group (n = 23, self-reported experience over 3 points) had an overall mean change count of 5.07 (SD = 1.85) and mean search time of 25.15 seconds (SD = 7.72). For the inexperienced group (n = 23, self-reported experience equal or less to 3 points) the averages were 4.17 (SD = 1.41) and 30.62 seconds (SD = 6.25), respectively. The difference in the change count was not statistically significant (p = 0.07), but there was a significant difference in the average values between the search times (t = -2.64, p = 0.01. d = 0.78). As the experienced group consisted mostly (82.6%) of male participants the difference between the experienced and inexperienced groups most likely also explains the differences between males and females reported above.

The experimental apparatus prohibited some subjects to wear their prescription glasses during the experiment. To see if this affected the study results, the average proportion of right answers over the 12 rooms was calculated for the group who had to remove the glasses (n = 11). The mean proportion was 0.83 (SD = 0.16). The average for the control group was 0.72 (SD = 0.17). Removing the glasses did not negatively affect the experimental performance.

Overall the subjects reported a low level of nausea or dizziness after the experiment (M = 1.2, SD = 0.6, on a scale from 1-5).

## EXPERIMENT 2

In the second experiment we collected preliminary online data by distributing the experimental program as a self-contained program on various websites related to virtual reality. The aim was to validate the feasibility of an online VR experiment.

### Materials and methods

### Participants

The online program was introduced as a change blindness experiment that anyone with the recommended hardware can participate. The program was distributed on four websites related to virtual reality - wearvr.com, riftcat.com, forums.oculus.com and vnslab.mozello.com. The number of downloads exceeded 300 instances and data from 25 study subjects (mean age 25.7 years, SD = 9.1, two female) were collected. The sample collection process ran for two months. Participants received feedback on their performance after the experiment.

### Stimuli and procedure

The stimuli was identical to that of experiment 1, with the exclusion of two rooms in the experimental block to make the study shorter in duration. We excluded one room from the foreground condition and one from the background condition (see figure S1 in the Supplemental Material available online). The background questionnaire was also much shorter, asking only the participants age, gender, experience with computer games and use of mnemonic techniques during the study. Instructions were given as an audio recording and through on-screen text. The experiment lasted approximately 15 minutes. The study protocol complied with the declaration of Helsinki, participation was anonymous, voluntary and the experiment could be stopped at any time.

### Apparatus

The 3D environments were constructed similarly to experiment 1. All participants were asked to use the Oculus Rift Development Kit 2 virtual reality headset (Oculus VR, LLC) to conduct the experiment at their own time and in a quiet environment, using headphones for noise isolation and experimental instructions. The instructions required participants to confirm that their system was capable of running the experience at 75 frames per second. For anonymity reasons no data was collected on the PC specifications used for the experiment.

**Data cleaning and analysis**

Data cleaning was similar to that of experiment 1. Of the total subject pool of 25 participants, 22 subjects were included in the final experimental data analysis. Three subjects were excluded due to a high number (>2) of answering errors caused probably by misinterpretations of the instructions.

**Results**

Experiment 2 was conducted online with a sample of 25 study subjects. Subjects completely failed to identify or misidentified the changed objects in 28 of the 220 trials (17.3%). For all the successful trials, the mean search time was 20.9 seconds (SD = 13.8) and on average, the changing object changed states 6.8 times (SD = 5) before being spotted. 5 study subjects out of 22 (22.7%) managed to successfully spot all the 10 changes presented in the experiment. The average self-reported previous experience with computer games was 4.2 (SD = 0.9). Subjects reported a relatively low level of nausea or dizziness after the experiment (M = 1,8, SD = 1, on a scale from 1-5). Six participants reported using mnemonic techniques to complete the task.

First the number of errors (false answer or out of time trials) was examined between conditions from the sample of 220 trials. On the proportion of trials where the person detected a change, the foreground condition yielded a mean value of 0.89 (SD = 0.13) and the background condition 0.76 (SD = 0.24). A two-tailed paired t-test assuming unequal variances revealed a statistically significant difference between the proportions of successful trials (t = 2.45, p = 0.02. d=0.66). Changing objects farther away from the subject were again detected less successfully. All following analysis were conducted without the data from the error trials.

When comparing the amount of changes needed for successful detection on 220 trials, no significant difference between conditions was found using Wilcoxon signed rank test with continuity correction (V = 106.5, p = 0.53. d = 0.18). The mean amount of changes for the foreground and background conditions were 6.57 (SD = 2.9) and 7.1 (SD = 3), respectively.

The mean search time for the foreground condition was 20.83 seconds (SD = 7.5) and for the background condition 23.12 seconds (SD = 11.6). However, the difference between the means was not statistically significant (t = -1.11, p = 0.28. d=0.23). Results between conditions are shown on figure 2 B.

We also compared the overall results of the offline and online samples (experiment 1 and 2, respectively). For maximum similarity we removed the same two extra rooms from the offline version that were absent from the online study when doing the analysis. There was no statistically significant difference between the proportions of correct answers, but there was a trend showing that in the online sample subjects detected more changes (t = 1.86, p = 0.07. d = 0.4). However there were significant differences in the amount of changes needed for successful change detection (t = 2.14, p = 0.04. d = 0.6) and in search time (t = -3.04, p = 0.004. d = 0.12). In the online sample, subjects needed less change cycles and were quicker to spot the changes. The results can also be seen on figure 3.

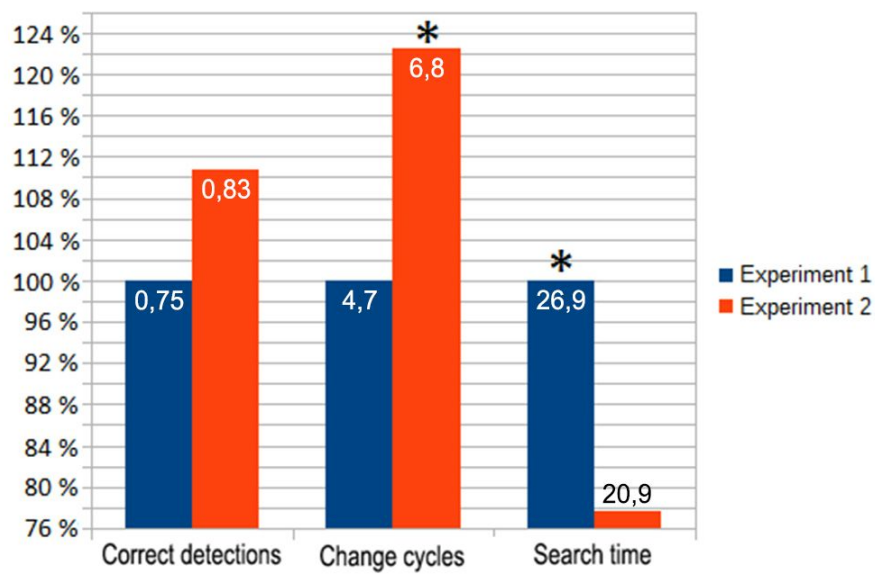

Figure 3. Result comparisons for Experiment 1 and Experiment 2 as percentage of Experiment 1 values (y axis). Numbers on the bars show absolute values. Asterisks denote statistically significant differences (p < 0.05).

**General discussion**

A quasi-realistic VR approach with a novel change induction paradigm was used to investigate change blindness in a more natural setting. A high amount of missed changes (nearly ¼ of all trials) was observed, as previous literature with VR setup has also shown (Steinicke et al, 2010; Suma et al, 2011). However, going further than these previous works, the present results indicate that the change blindness effect persists even when the subject can freely look around in the environment (in contrast to Steinicke et al, where the scene was static) and is explicitly instructed to search for changes (in contrast to Suma et al, where the subjects were naïve). Judging by the proportion of correct identifications and number of changes needed for detection, the data analysis confirmed previous results with 2D display setups, with foreground changes being significantly easier to detect (Mazza et al, 2005; Turatto et al, 2002) in both experiments.

There was no significant difference in search time between the foreground and background conditions. Previous research on this matter is sparse, as neither Mazza et al (2005) nor Turatto et al (2002) analyzed temporal data in their foreground/background paradigms. One explanation for the lack of difference could be that the dimensions of the virtual rooms used in the present study (approximately 10 meters wide and 6 meters long) were too small to generate noticeable time differences when switching attention between the foreground and background depth planes. The result could also be attributed to our paradigm that relied on subjects head movements to induce the change in the scene. From the results we know that subjects approached the task in different ways.

Since the method used for introducing the changes in the scene was novel, background information about different pre-defined strategies was collected after the experiment. The most popular strategies were checking the objects one-by-one or turning one's head repeatedly to detect the changes, used by slightly over half of the participants. Repeated head movements had significant effects on the results, with much shorter search times in the foreground condition. However, this improvement came with the cost of more changes being needed to detect the target object. It is worth pointing out that this kind of repeated head turning produces similar visual effects to the "flicker paradigm" used widely in many two-dimensional change blindness studies (Simons & Rensink, 2005). For more natural results, future studies may want to prevent this kind of strategy use by prohibiting it in the instructions or introducing a minimum interval between the changes to render the strategy obsolete.

According to verbal comments, sometimes inattentional blindness occurred, when searching for small changes and therefore missing big ones. Many subjects who failed to detect big changes in a given room were genuinely surprised when the change was revealed after the time ran out. One participant commented: "I can't believe how difficult it was to remember what was in the room." This "looking without seeing" phenomena has been previously explained by the inconceivable nature of such changes in real situation, that cannot be integrated into the subjects momentary conceptual framework (O'Regan et al, 2000). It could be that the virtual environment was used as an external memory to be probed when details need to be obtained, as has been suggested previously (O'Regan et al, 2000).

Although the online experiment had a low number of participants, some observations can be made about the sample in comparison to the lab condition. The age of online participants was generally greater and mainly experienced male computer gamers participated. This can be explained by the current state of consumer VR technology that is mostly targeted to software developers. As the market expands and VR finds its way into more households, a more representative sample can be expected. The results from the online sample resembled those from the experienced gamer group in the offline protocol. A reason for the low number of online participants compared to downloads can be various software glitches, as the program may crash on unexpected hardware setups. Also, even though the online experiment lasted about 15 minutes, this could have been too long and repetitive for potential subjects. The sheer fact that the 25 subject sent in data from anywhere in the world holds great promise for the future of online VR experiments. One promising option to recruit more participants is making the experiments more game-like, as has been also proposed elsewhere (Scarfe & Glennerster, 2015).

**Conclusion**

Using VR with a novel change induction paradigm allows for a natural paradigm to study human attention. It was observed that subjects often miss the relatively large changes or take a long time to spot them. Changes in the foreground were detected more easily than changes in the background, confirming previous results and suggesting that attention is directed more towards foreground objects (Mazza et al, 2005; Turatto et al, 2002). Further studies should explore the effects of longer distances or different environments on change blindness performance. Results with an online paradigm showed the possibility to conduct reliable and large-scale VR field studies already in the near future.

**Author Contributions**

M. Vasser developed the study concept and performed data collection. M. Kängsepp wrote the scripts for the program used in the study. All authors contributed to the study design and testing. M. Vasser performed the data analysis and interpretation under the supervision of J. Aru. M. Vasser drafted the manuscript, and M. Kängsepp and J. Aru provided critical revisions. All authors approved the final version of the manuscript for submission.

**Acknowledgments**

The authors like to thank Renate Rutiku, all the study subjects and OÜ Psühhobuss.

**Competing interests**

The authors declare that they have no competing interests.

**Figure S1**

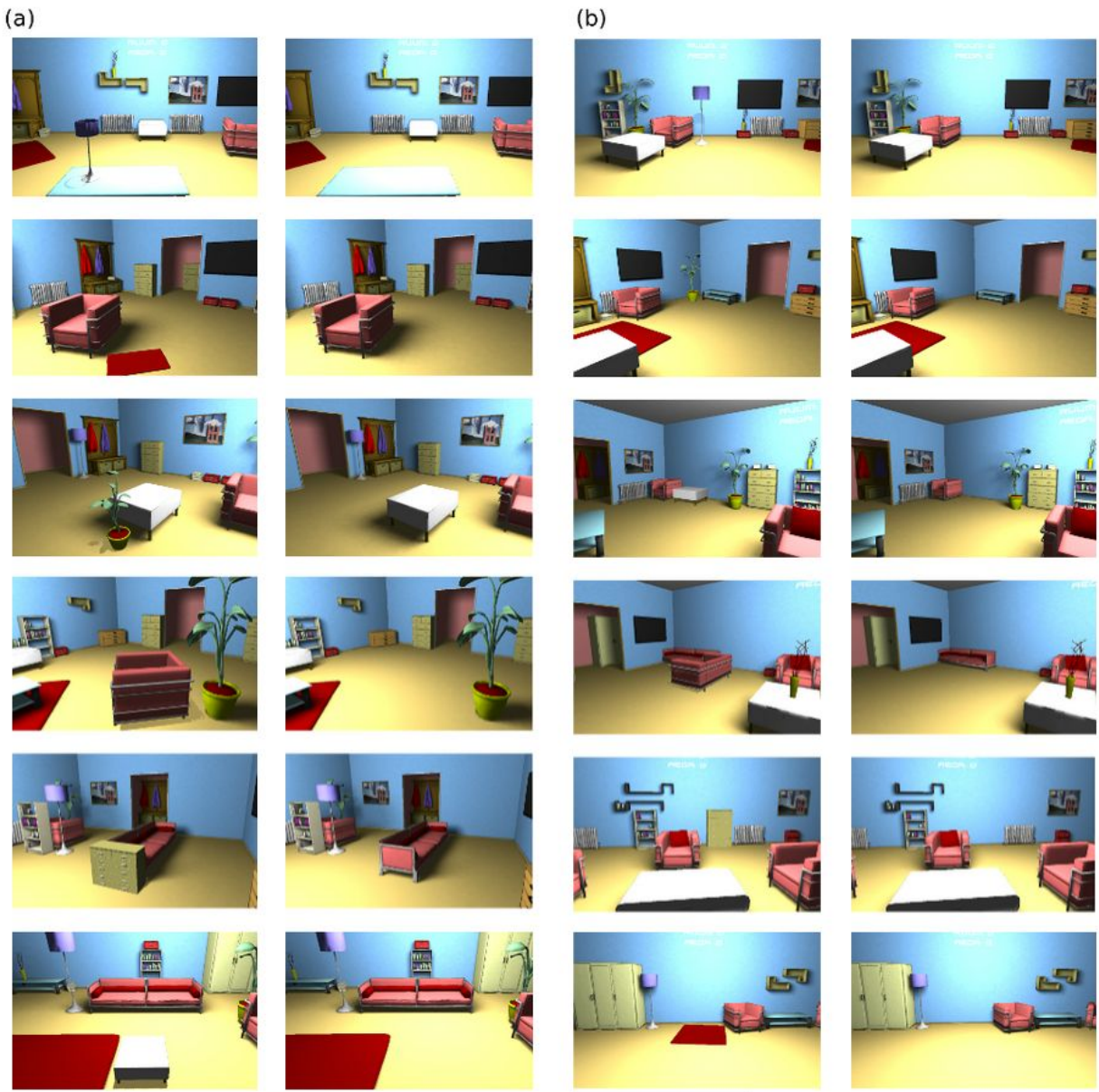

Figure S1. All the changes presented in experiment 1. Foreground condition is shown in block (a), background condition in block (b). For experiment 2, the last change from both conditions was omitted (foreground table and background carpet).